\documentclass[12pt]{iopart}
\usepackage{graphicx}
\begin{document}
% Journal identifier can be put here if required, e.g.
%\jl{14}

\title[Superquasicrystals]
{Superquasicrystals: selfsimilar ordered structures with non-crystallographic point symmetries}

\author{Komajiro Niizeki\
\footnote[3]{niizeki@cmpt.phys.tohoku.ac.jp}
 and Nobuhisa Fujita\dag
}

\address{\ Department of Physics, Graduate School of Science, Tohoku University, Sendai 980-8578, Japan}

\address{\dag\ Structural Chemistry, Arrhenius Laboratory, Stockholm University, 10691 Stockholm, Sweden}

\begin{abstract}
We present a systematic method of constructing limit-quasiperiodic structures 
with non-crystallographic point symmetries. 
Such structures are different aperiodic ordered structures from quasicrystals, 
and we call them ``superquasicrystals". 
They are sections of higher-dimensional limit-periodic structures constructed on "super-Bravais-lattices". 
We enumerate important super-Bravais-lattices. 
Superquasicrystals with strong selfsimilarities form an important subclass. 
A simplest example is a two-dimensional octagonal superquasicrystal. 
\end{abstract}



%\maketitle
%\date{\today}

%-------------------------------------------------------

Quasicrystals (QCs) as well as the Penrose patterns are aperiodic ordered structures 
having not only long-range positional orders 
but also non-crystallographic point symmetries (NCPSs) \cite{SBGC, P74}. 
Moreover, they have self-similarities, and are produced by substitution rules
(equivalently, inflation rules).
They can be described alternatively as quasiperiodic point sets, 
given as sections of higher-dimensional crystals, or {\it hypercrystals} \cite{LeSt, KaDu, Bak}.
From the viewpoint of aperiodic ordered structures, however,
there can be another type of selfsimilar structures with long-range positional orders, namely,
{\it limit-quasiperiodic structures} (LQPSs), 
which are sections of higher-dimensional limit periodic structures \cite{GK}. 
An LQPS is usually given as the set of vertices of an aperiodic tiling generated 
by a substitution rule with the {\it Pisot property} \cite{GK, LGJJ, K93}. 
On the other hand, a limit periodic structure and LQPSs are incorporated in {\it model sets}, 
which are projections of some transcendental host structures 
(see \cite{BMS98} and references cited therein). 
Since a limit-quasiperiodic structure is not periodic, it as well as a QC allows any NCPS, 
and we shall call such a structure a ``superquasicrystal" (SQC). 
An SQC is a section of a higher-dimensional limit periodic structure, 
which we shall call a {\it supercrystal}; 
we must distinguish the supercrystal from a hypercrystal. 
Regrettably, no concrete example of SQCs has been known before. 
In this Letter, the following four points will be discussed:
i) A {\it super-Bravais-lattice}, which is a geometric object in a Euclidean space, 
is introduced as a host structure on which supercrystals are constructed. 
ii) Important super-Bravais-lattices with NCPSs are enumerated. 
iii) It is shown that an SQC can have a self-similarity characterized by a Pisot number 
which is severely restricted by the super-Bravais-lattice. 
iv) As a simplest example, an octagonal SQC is presented. 

We begin with a brief review of QCs (for a finer review, see \cite{Ya96}). 
The point groups which are important in application to physical QCs include
${\rm D_8}$ (octagonal), ${\rm D_{10}}$ (decagonal) 
and ${\rm D_{12}}$ (dodecagonal) in two dimensions (2D) and ${\rm Y_h}$ (icosahedral) in 3D; 
all these point groups are isometric. 
A QC having one of these point groups is a $d$-dimensional 
section of a hypercrystal in the $2d$-dimensional space $E_{2d}$ with $d=2$ or $3$ 
being the number of dimensions of the physical space $E_d$. 
The entities of the hypercrystal are not atoms but geometric objects 
called {\it hyperatoms} (atomic surfaces, windows, etc.), 
which are bounded open domains in the internal space, $E_d^\perp$, 
i.e., the orthogonal complement to $E_d$. 
We consider a hypercrystal constructed by locating one kind of hyperatoms 
onto the sites of a Bravais lattice ${\hat \Lambda}$ in $E_{2d}$. 
There exists only one Bravais lattice for each of the three 2D point groups, 
but there exist three types of icosahedral Bravais lattices, 
namely, primitive, face-centered and body-centered types. 
An infinite number of different QCs are obtained from a single hypercrystal by 
choosing different sections which are parallel to $E_d$, 
and they form a so-called {\it locally isomorphic class} (LI-class) \cite{LeSt}. 
The point group ${\hat G}$ of ${\hat \Lambda}$ is isomorphic 
with a noncrystallographic point group $G$ in $d$ dimensions. 
More precisely, ${\hat G} = G \oplus G^\perp$, where $G$ and $G^\perp \,(\simeq G)$
are groups acting onto $E_d$ and $E_d^\perp$, respectively.
Let $\pi$ and $\pi^\perp$ be the projectors projecting $E_{2d}$ onto $E_d$ and $E_d^\perp$, respectively.
Then, $\Lambda := \pi{\hat \Lambda}$ and $\Lambda^\perp := \pi^\perp{\hat \Lambda}$ are 
modules with $d$ generators, and are dense sets. 
They have a scaling symmetry whose scale is a quadratic irrational, $\tau$, 
where $\tau = 1 + \sqrt{2}$, $\frac{1}{2}(1 + \sqrt{5}\,)$ or $2 + \sqrt{3}\,$ 
for the case of the three 2D point groups, ${\rm D_8}$, ${\rm D_{10}}$ or ${\rm D_{12}}$, respectively, 
while $\tau = 2 + \sqrt{5}\,$ for the primitive-icosahedral case 
but $\tau = (1 + \sqrt{5}\,)/2$ for the other two icosahedral cases; 
$\tau$ is a Pisot number and also a unit in ${\bf Z}[\tau] := \{n + m\tau \,|\, n, \,m \in {\bf Z}\}$. 
Let $\varphi$ be the scaling operation with the ratio $\tau$. 
Then, any member of the infinite group, ${\cal G} := \langle G, \, \varphi\rangle$, 
generated by $G$ and $\varphi$ is an automorphism of $\Lambda$. 
A QC is ``point diffractive" in the sense that 
its structure factor is composed only of Bragg spots.
The position vectors of the Bragg spots form the Fourier module, 
which corresponds to the reciprocal lattice for a periodic crystal.
The Fourier module is given by $\Lambda^* := \pi{\hat \Lambda}^* \subset E_d^*$ with 
${\hat \Lambda}^*$ being the reciprocal lattice of ${\hat \Lambda}$ 
and $E_d^*$ the dual space to $E_d$. 

We have already seen several examples of triads of objects, 
$\{X, \, X^\perp, \,{\hat X}\}$, associated with the three worlds, $E_d$, $E_d^\perp$
and $E_{2d}$, where $X = \pi{\hat X}$ and $X^\perp := \pi^\perp{\hat X}$. 
Any member of a triad uniquely determines the remaining two; 
in particular, ${\hat X} = \pi^{-1}X$ is the lifted version of $X$. 
Many relationships are isomorphic among the three worlds, 
and a relationship in one of the three worlds can be readily 
translated to those in the other two \cite{KaDu}; 
if $X$ and $X^\perp$ above are sets, the symbol ``$\perp$" turns out 
a bijection from $X$ onto $X^\perp$. 
There exists an important triad of linear maps
$\{\varphi, \, \varphi^\perp, \,{\hat \varphi}\}$ with $\varphi^\perp$ 
being a scaling with the ratio ${\bar \tau}$, the algebraic conjugate of $\tau$, whereas 
${\hat \varphi} = \varphi \oplus \varphi^\perp$is an automorphism of ${\hat \Lambda}$: 
${\hat \varphi}{\hat \Lambda} = {\hat \Lambda}$. 
It follows that a QC has a self-similarity with ratio $\tau$ \cite{KaDu, Ni89a}. 

Another important triad of linear maps is associated with a bi-similarity transformation 
${\hat \sigma} = \sigma \oplus \sigma^\perp$, 
where $\sigma$ and $\sigma^\perp$ are similarity transformations acting onto $E_d$ and $E_d^\perp$, respectively, 
while ${\hat \sigma}$ satisfies ${\hat \Lambda}_1 := {\hat \sigma}{\hat \Lambda} \subset {\hat \Lambda}$. 
${\hat \Lambda}_1$ is the last member of the triads $\{\Lambda_1, \, \Lambda_1^\perp, \,{\hat \Lambda}_1\}$, 
in which $\Lambda_1 = \sigma\Lambda$ and $\Lambda_1^\perp = \sigma^\perp\Lambda^\perp$ 
are {\it similar submodules }(SSMs) of $\Lambda$ and $\Lambda^\perp$, respectively. 
Moreover, $\Lambda_1$ as well as $\Lambda$ is invariant against the action of ${\cal G}$. 
We shall denote by $m$ the index of $\Lambda_1$ in $\Lambda$: $m := [\Lambda: \Lambda_1]$.  
The scaling transformation $\varphi$ is a special similarity transformation since it is invertible 
and, consequently, $m = 1$. 
However, if $m > 1$, $\Lambda$ is a larger set than $\Lambda_1$, 
and hence $\sigma$ is not invertible in $\Lambda$.
${\hat \Lambda}_1$ in this case is a sublattice of ${\hat \Lambda}$ or, 
in the terminology of crystallography, a superlattice of ${\hat \Lambda}$, 
and is divided into $m$ equivalent sublattices to ${\hat \Lambda}_1$. 
We shall call ${\hat \Lambda}_1$ a quasi-similar superlattice of ${\hat \Lambda}$. 
The similarity transformation $\sigma$ combining an SSM, $\Lambda_1$, 
with $\Lambda$ by $\Lambda_1 = \sigma\Lambda$ is, however, not uniquely determined by $\Lambda_1$ 
because any member of the set ${\cal G}(\Lambda_1) := \sigma{\cal G} = {\cal G}\sigma$ 
satisfies the same condition. 
The set of all the similarity ratios of the members of ${\cal G}(\Lambda_1)$ 
is given by $\{|\sigma|\tau^k \,|\, k \in {\bf Z}\}$. 
Let ${\hat {\cal B}}$ be the set of all the quasi-similar superlattices of ${\hat \Lambda}$. 
Then, it is a member of a triad, $\{{\cal B}, \, {\cal B}^\perp, \,{\hat {\cal B}}\}$, 
and ${\cal B}$ is the denumerable set of all the SSMs of $\Lambda$. 
The set, $\Sigma$, of all the similarity transformations associated with ${\cal B}$ form a semigroup. 
We may expect that there exists a bijection between ${\cal B}$ and the quotient semigroup, $\Sigma/{\cal G}$. 
This is true provided that, for the case of the 2D point group ${\rm D_{12}}$, 
${\cal G}$ is slightly modified as made in the next paragraph. 

Prior to proceeding to the subject of super-Bravais-lattices, 
we shall investigate in more detail SSMs of $\Lambda$ 
for the case of the 2D point group ${\rm D_p}$ with $p = 8$, 10 or 12. 
An important member of ${\cal B}$ is the one written as $\sigma_p = |\sigma_p|\rho_{2p}$ 
with $|\sigma_p| := 2\cos{(\pi/p)}$ and $\rho_k$ being 
the rotation operation through $2\pi/k$ \cite{Ni89a, Ni89b}. 
%where $|*|$ stands for the scale of the similarity transformation $*$. 
The index $m_p$ of the SSM, $\sigma_p\Lambda$, is equal to two, five or one 
for $p = 8$, 10 or 12, respectively. 
Since $\sigma_{12}$ is invertible and $|\sigma_{12}| = \sqrt{\tau\,}$, 
we have to redefine the map $\varphi$ for $p = 12$ by $\sigma_{12}$ 
and, correspondingly, the automorphism group ${\cal G}$ of $\Lambda$ is redefined. 
There exist two types of SSMs: we call $\Lambda_1 = \sigma\Lambda$ a type I or II SSM 
if $\sigma$ is chosen to be a simple scaling or not, respectively. 
A complete discussion for possible SSMs has been made in \cite{Ni89b}. 
A type I SSM is written with a positive number $\nu \in {\bf Z}[\tau]$ as $\nu\Lambda$, 
and its index is given by $m = [N(\nu)]^2$ with $N(\nu) := \nu{\bar \nu}$. 
A simplest SSM for $p = 12$, for example, is given by $(1 + \sqrt{3}\,)\Lambda$, 
whose index is equal to four. 
On the other hand, type II SSMs are somewhat complicated. 
We shall call a type II SSM proper 
if the point group ${\rm D_p}$ of $\Lambda$ leaves it invariant. 
If it is not proper, the common point group between it and $\Lambda$ is equal to ${\rm C_p}$. 
In this report, we shall ignore ``improper" type II SSMs (cf. \cite{Ni89a}). 
Then, there exist no type II SSMs for $p = 12$. 
A simplest type II SSM for $p = 8$ or 10 is $\sigma_p\Lambda$. 
A general type II SSM is written with $\nu \in {\bf Z}[\tau]$ as 
$\nu\sigma_p\Lambda$ and $m = m_p[N(\nu)]^2$. 
We may write $\sigma := \nu\sigma_p = |\sigma|\rho_{2p}$ with $|\sigma| = \nu|\sigma_p|$. 
Then, $\sigma^2\Lambda$ is identical to the type I SSM, $|\sigma|^2\Lambda$, 
because $(\rho_{2p})^2 \,(= \rho_p) \in {\rm D_p}$. 
It follows that $|\sigma|^2 \in {\bf Z}[\tau]$. 
In particular, $|\sigma_8|^2 = 2 + \sqrt{2\,} = \sqrt{2\,}\,\tau$, and 
$\sigma_8^2\Lambda = \sqrt{2\,}\,\Lambda$ for $p = 8$ because $\tau\Lambda = \Lambda$. 
Similarly, $|\sigma_{10}|^2 = \sqrt{5\,}\,\tau$ and $\sigma_{10}^2\Lambda = \sqrt{5\,}\,\Lambda$ for $p = 10$. 
Note that $|\sigma_p| \notin {\bf Z}[\tau]$. 

The modules $\Lambda_n := \sigma^n\Lambda$, $\forall n \in {\bf N}$, satisfy 
$\Lambda \supset \Lambda_1 \supset \Lambda_2 \supset \cdots $ and $[\Lambda: \Lambda_n] = m^n$. 
We shall denote by ``$\stackrel{n}{\equiv}$" the equivalence relationship 
introduced into $\Lambda$ by the residue module, $\Lambda/\Lambda_n$. 
Its important property is the following: if $\ell \stackrel{n}{\equiv} \ell'$ for $\ell, \,\ell' \in \Lambda$, 
$\ell \stackrel{n'}{\equiv} \ell', \,\forall n' \le n$. 
As a consequence, $\Lambda$ becomes a normed module 
if a non archimedean norm of $\ell \in \,\Lambda$ is defined by $||\ell|| := 1/2^n$ 
with $n$ being the largest number satisfying $\ell \stackrel{n}{\equiv} 0$. 
It is just a metric space called an {\it inverse system}; 
different vectors in $\Lambda$ are ``coloured" by different colours, 
and $\Lambda$ is regarded as a ``coloured module" with an infinite number of colours. 
This structure can be transferred to ${\hat \Lambda}$, 
yielding a $2d$-dimensional coloured lattice, ${\cal L}$. 
We shall call ${\cal L}$ a {\it super-Bravais-lattice} 
because different supercrystals are constructed on it as shown shortly. 
Since ${\hat \Lambda}_n := \pi^{-1}\Lambda_n$ is superlattice of ${\hat \Lambda}$, 
${\cal L}$ is, intuitively, a recursive superlattice structure. 
The map $\sigma$ as well as the group ${\cal G}$ acts naturally onto ${\cal L}$, 
and $\sigma{\cal L} \subset {\cal L}$. 
We shall turn our arguments to the dual space (or the reciprocal space). 
The modules $\Lambda_n^* := \sigma^{-n}\Lambda^*, \, \forall n \in {\bf Z}$, 
satisfy $\Lambda_n^* = \sigma\Lambda_{n+1}^* \subset \Lambda_{n+1}^*$. 
The denumerable set ${\hat \Lambda}_\infty^* := \pi^{-1}\Lambda_\infty^*$ with 
$\Lambda_\infty^* := \Lambda_0^* \cup \Lambda_1^* \cup \Lambda_2^* \cup \cdots$ 
is the dual module to ${\cal L}$. 
It is not finitely generated in contrast to $\Lambda^*$. 
It is invariant against the action of the infinite group, $\langle G, \, \sigma\rangle$. 
We can divide $\Lambda_\infty^*$ into the disjoint sets, 
$\Delta_n := \Lambda_n^* - \Lambda_{n-1}^*, \, n \in {\bf Z}$, which are not modules. 
They are invariant against the action of $G$ and 
satisfy $\Delta_n = \sigma\Delta_{n+1}$ and $\Delta_n \subset \Lambda_n^*$. 
We may write as $\Lambda_\infty^* = \Lambda_0^* + \Delta_1 + \Delta_2 + \cdots$ 
because $\Lambda_0^*$ is disjoint with $\Delta_n, \,n > 0$. 
Any vector in $\Delta_n$ is indexed with the generators of $\Lambda_n^*$ by $2d$-integers. 

In order to construct a supercrystal on the super-Bravais-lattice, ${\cal L}$, 
we need a set ${\cal A}$ consisting of all the allowed hyperatoms, 
whose diameters are assumed to be bounded by a positive number. 
We can identify a hyperatom with its characteristic function on $E_d^\perp$, 
and ${\cal A}$ is embedded into $L^1$, the function space with the $p = 1$ norm. 
Let $\alpha$ be a map from $\Lambda$ into ${\cal A}$. 
Then, a supercrystal is specified by ${\cal L}$ and $\alpha$; 
we shall denote it by ${\cal S}({\cal L}, \,\alpha)$. 
The hyperatoms in a supercrystal are {\it not uniform} but 
their shapes, sizes, and/or orientations are determined by the colours of the relevant sites. 
We assume $\alpha$ to be {\it uniformly-continuous} in the sense 
that, for any $\varepsilon > 0$, there exists an integer $n$ such that 
$||\alpha_\ell - \alpha_{\ell'}||_1 < \varepsilon, \; \forall \ell, \,\ell' \in \,\Lambda$ 
with $\ell \stackrel{n}{\equiv} \ell'$, 
where $\alpha_\ell \in {\cal A}$ stands for the image of $\ell \in \Lambda$. 
The point group of the supercrystal is identical to ${\hat G}$ 
if $\alpha G = G^\perp \alpha$ and ${\cal A}$ is invariant against the action of $G^\perp$. 
The map fulfills these conditions if $\alpha_\ell$ is, for instance, 
a regular $p$-gon whose size is given by $f(||\ell||)$ with $f(x)$ being a continuos function 
bounded from both sides by two positive numbers. 
Let $\alpha$ be a special map satisfying $\alpha_\ell = \alpha_{\ell'}, 
\; \forall \ell, \,\ell' \in \,\Lambda$ with $\ell \stackrel{n}{\equiv} \ell'$. 
Then, ${\cal S}({\cal L}, \,\alpha)$ degenerates into a hypercrystal 
whose translational group is given by ${\hat \Lambda}_n$. 
If the {\it supremum norm} is introduced into the ``function space" of maps, $\{\alpha\}$, 
a supercrystal can be approximated in any precision in this norm by a hypercrystal 
to be called an {\it approximant hypercrystal}. 
This means that the supercrystal is limit periodic \cite{GK}. 

An SQC obtained from the supercrystal, ${\cal S}({\cal L}, \,\alpha)$, 
is parametrized by the phase vector, $\phi \in E_d^\perp$ 
specifying the ``level" at which the section of the supercrystal is taken. 
It is represented as $S({\cal L}, \,\alpha, \,\phi) := {\cal S}({\cal L}, \,\alpha) \cap (\phi + E_d)$, 
which is a disctrete subset of $\Lambda$. 
If identical point scatters with the unit scattering strength are located on the sites of the SQC, 
we obtain a scatterer field on $E_d$: 
\begin{eqnarray}
s({\bf x}) = \sum_{\ell \in \,\Lambda}\;\alpha_\ell(\phi - \ell^\perp)\delta({\bf x} - \ell)
\label{scatter}\end{eqnarray}
with $\alpha_\ell({\bf x}^\perp)$ being the characteristic function of $\alpha_\ell$. 
This form of the scatterer field can be generalized for the case 
where $\alpha$ is a generic uniformly-continuous map from $\Lambda$ into $L^1$; 
the scattering strength of a site may now depend on the ``colour" of the site. 
This generalized scatterer field is point diffractive because 
it can be approximated in any precision in the supremum norm 
by one of {\it its quasiperiodic approximant}. 
Thus, an SQC is a perfectly ordered structure with a long-range order.
An SQC is, in fact, a model set. 
This is shown by identifying $\alpha$ with the set 
$\{(\alpha_\ell, \,\ell) \,|\, \ell \in \,\Lambda\} \,(\subset E_d^\perp\times\Lambda)$, 
which is basically a {\it window} in the theory of the model set \cite{BMS98}. 

The Fourier module of the SQC is given by $\Lambda_\infty^*$, 
and each Bragg spot is indexed by $(2d+1)$-integers; 
the last index specifies the superlattice order $n$. 
We should emphasise that the Fourier module of 
$S({\cal L}, \,\alpha, \,\phi)$ is determined solely by ${\cal L}$, the super-Bravais-lattice. 
The Fourier transform of the distribution Eq.\ref{scatter} is a distribution in $E_d^*$: 
\begin{eqnarray}
s^*({\bf Q}) = \sum_{\ell^* \in \,\Lambda_\infty^*}\;\alpha_{\ell^*}^*(- (\ell^*)^\perp)\delta({\bf Q} - \ell^*), 
\label{scatter*}\end{eqnarray}
where $\alpha^*$ stands for a map from $\Lambda_\infty^*$ into $L^1$ defined on $(E_d^*)^\perp$. 
Since ${\cal A} \subset L^1$, we can define an average over a set of hyperatoms. 
Then, for $n \in {\bf N}$, we can define naturally the $n$-th order averaged hypercrystal,
in which the hyperatom assigned to ${\hat \ell} \in {\hat \Lambda}_n$ is the average over all the hyperatoms 
on the residue class ${\hat \ell} + {\hat \Lambda}_n$. 
It can be shown readily that the Fourier transform of the scatterer field of the averaged QC 
is identical to the one obtained from Eq.\ref{scatter*} by 
restricting the summand to $\Lambda_n^*$. 
This allows us to determine the map  $\alpha^*$. 

An SQC has always a self-similarity in the sense that 
it includes a subset which is geometrically similar to itself. 
Its proof is basically similar to the one which was made in \cite{Ni89c} for the case of uniform hyperatoms. 
The key concept in the proof is {\it Pisot maps}. 
A similarity transformation $\sigma$ is called a Pisot map if
$|\sigma^\perp| < 1$; this implies $|\sigma| > 1$ because $m = (|\sigma|\,|\sigma^\perp|)^d \ge 1$.
The map, $\varphi$, is a Pisot map because $|\varphi^\perp| = \tau^{-1} < 1$. 
Since $\varphi{\cal G}(\Lambda_1) = {\cal G}(\Lambda_1)$, 
${\cal G}(\Lambda_1)$ includes an infinite number of Pisot maps. 
A similar subset of $S := S({\cal L}, \,\alpha, \,0)$ is written as
$S_1 := \sigma S$ with $\sigma \in {\cal G}(\Lambda_1)$ being a Pisot map
\cite{Ni89c}. 
We may say that the SQC, $S$, is strongly selfsimilar 
if it and its inflation, $S_1$, are mutually locally-derivable (MLD: for MLD see \cite{Ba91}). 
This is not necessarily the case for a generic $\alpha$. 
Regrettably, we have yet found no condition for $\alpha$ such that a strongly selfsimilar SQC is obtained. 
A weak self-similarity is of no physical interest as discussed in \cite{Ni89c}. 

Since $S_1 \subset \Lambda_1$, the inflated SQC 
come only from one of $m$ submodules into which $\Lambda$ is divided. 
More generally, the $n$-th inflation, $S_n := \sigma^nS$, is a subset of $\Lambda_n$. 
This is in sharp contrast to the case of a QC, for 
which the inflated QC come evenly from different submodules on account of $\varphi\Lambda = \Lambda$. 
The self-similarity ratio $|\sigma|$ (or its square $|\sigma|^2$) is a Pisot number in ${\bf Z}[\tau]$ 
if the map $\sigma$ is of the type I (or II). 
However, it is not a unit in contrast to the self-similarity ratio of a QC. 
In particular, the smallest value of $|\sigma|$ is given by $|\sigma_8|$, 
$|\sigma_{10}|\tau$ or $1 + \sqrt{3}\,$ for the case of the 2D point group 
${\rm D_p}$ with $p = 8$, 10 or 12, respectively. 
It should be emphasised that self-similarity of an SQC as well as a QC is 
a natural consequence of its NCPS. 

A limit-quasiperiodic tiling produced by a substitution rule is necessarily strongly selfsimilar. 
Properties of 2D and 3D tilings of this type have been extensively investigated in \cite{GK}, 
and a recursion formula determining the Fourier transform of the relevant scatterer field is presented. 
A limit-quasiperiodic tiling is considered to be an SQC only if it has an NCPS. 
However, such a tiling has not been known so far \cite{note}.

\begin{figure}[hbt]
\begin{center}\includegraphics[width = 8.5cm]{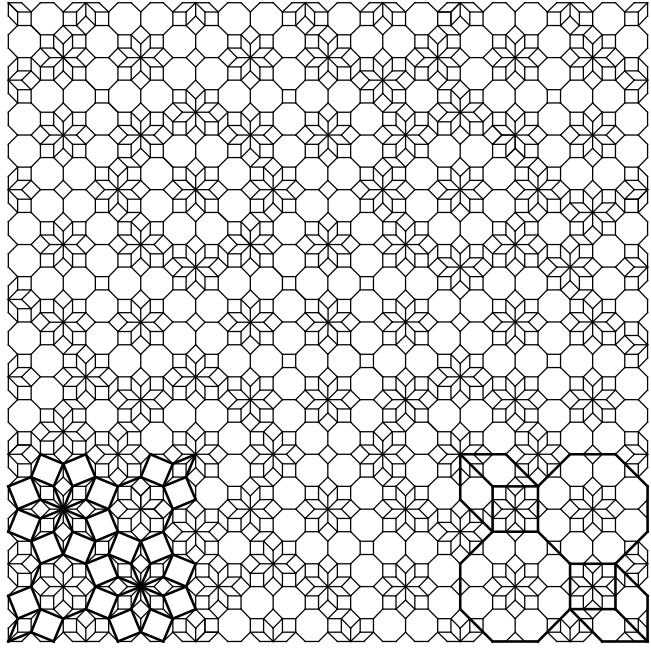} 
\caption{An octagonal SQC, $S$, associated with a tiling produced by a substitution rule for the three kinds of tiles. 
The left-bottom corner is the center of the exact octagonal symmetry. 
This center is the origin of $E_2$.
A part of its inflation, $\sigma_8S$, is shown with thick lines at the left-bottom, 
while a part of the double inflation, $(\sigma_8)^2S = (2 + \sqrt{2\,})S$, is at the right-bottom. 
} \label{Fig.1}
\end{center}\end{figure}

We have discovered a few examples of substitution rules which generate SQCs.
The simplest of them is the octagonal SQC in Fig.\ref{Fig.1} (cf. \cite{WIS}).
The relevant SSM is given by $\sigma_8\Lambda$.
We have determined yet the exact form of the relevant map, $\alpha$, of this SQC. 
Our preliminary investigation strongly indicates, however, 
that the hyperatoms are topologically discs with fractal boundaries. 
Strong Bragg spots of the SQC are located on $\Lambda^*$, the Fourier module of an octagonal QC, 
while those on $\Delta_n, \,n > 0$ are all weak; 
the latters are superlattice reflections in the terminology of crystallography. 
Interestingly, we have noticed that the distribution of Bragg spots 
exhibits a pattern strongly reflecting the symmetry of $\sigma_8$. 

The section, $S_{\rm s}$, of the octagonal SQC in Fig.\ref{Fig.1} 
through the horizontal line at the bottom yields 
a 1D limit-quasiperiodic tiling with two intervals, $S$ and $L$, 
whose lengths satisfy $|L|/|S| = \sqrt{2\,}$. 
The tiling is produced by the substitution rule: $S \to SLS$, $L \to LSSL$. 
It is, alternatively, given as a section of a 2D limit-periodic structure 
which is a 2D section of the relevant supercrystal, ${\cal S}$. 
On the other hand, the projection, $S_{\rm p}$, of the octagonal SQC onto the same line 
is identical to $S_{\rm s}/\sqrt{2\,}$, 
which is a section of a 2D limit-periodic structure given as a 2D projection of ${\cal S}$. 
Conversely, a 2D SQC can be constructed by the grid method from a 1D LQPSs. 
It is readily shown that the relevant hyperatoms are polygons. 

The present theory is readily extended to the case of the icosahedral point group in 3D.
Only the type I SSMs concern this case, and the index of an icosahedral SSM, 
$\nu\Lambda$, is given by $m = [N(\nu)]^3$. 

There exists a bijection between the infinite set, ${\cal B}$, 
and the set of all the super-Bravais-lattices on a single Bravais lattice, ${\hat \Lambda}$, 
which is the host lattice of QCs. 
This means that the world of SQCs is far richer than that of QCs, 
which is contrary to a previous conception. 
SQCs together with QCs form an important class of aperiodic ordered structures 
with non-crystallographic point symmetries. 

The authors are grateful to O. Terasaki.


\begin{thebibliography}{50}
\bibitem{note} A tiling in \cite{WIS} is approximately an SQC but 
its ``octagonal symmetry" is broken as pointed out in \cite{Ni89c}.
\end{thebibliography}
\end{document}